# Covert Wireless Communication with a Poisson Field of Interferers

Biao He, *Member, IEEE,* Shihao Yan, *Member, IEEE,* Xiangyun Zhou, *Senior Member, IEEE,* Hamid Jafarkhani, *Fellow, IEEE*

## Abstract

In this paper, we study covert communication in wireless networks consisting of a transmitter, Alice, an intended receiver, Bob, a warden, Willie, and a Poisson field of interferers. Bob and Willie are subject to uncertain shot noise due to the ambient signals from interferers in the network. With the aid of stochastic geometry, we analyze the throughput of the covert communication between Alice and Bob subject to given requirements on the covertness against Willie and the reliability of decoding at Bob. We consider non-fading and fading channels. We analytically obtain interesting findings on the impacts of the density and the transmit power of the concurrent interferers on the covert throughput. That is, the density and the transmit power of the interferers have no impact on the covert throughput as long as the network stays in the interference-limited regime, for both the non-fading and the fading cases. When the interference is sufficiently small and comparable with the receiver noise, the covert throughput increases as the density or the transmit power of the concurrent interferers increases.

## Index Terms

Physical layer security, covert wireless communication, stochastic geometry, signal processing.

## I. Introduction

As an unprecedented amount of private and sensitive information is transmitted over the open wireless medium, secure communication against eavesdropping has drawn significant attention

The work of S. Yan and X. Zhou was supported under Australian Research Council's Discovery Projects funding scheme (DP150103905).

B. He and H. Jafarkhani are with the Center for Pervasive Communications and Computing, University of California at Irvine, Irvine, CA 92697, USA (email: {biao.he, hamidj}@uci.edu).

S. Yan and X. Zhou are with the Research School of Engineering, The Australian National University, Canberra, ACT 2601, Australia (e-mail: {shihao.yan, xiangyun.zhou}@anu.edu.au).

from wireless researchers and engineers [1, 2]. In certain circumstances, protecting the content of the transmitted information is not sufficient. In order to avoid monitoring and maintain privacy, the communication itself is sometimes required to stay undetectable in the presence of wardens, which motivates the increasing effort devoted to the study on covert communication. In principle, the techniques of physical layer security and covert communication are designed or studied to achieve different objectives. Physical layer security focuses on protecting the content of communications, while covert communication hopes to hide the communication process. Thus, it is usually not feasible to directly apply the studies and techniques of physical layer security to covert communication.

Although spread spectrum techniques have been studied since a century ago for hiding wireless transmissions, a recent interest in the fundamental limits of covert communication has arisen. Concretely, the studies in the recent trend are interested in the throughput at which the transmitter and the receiver may communicate reliably while guaranteeing a low probability of detection from the warden. In particular, a square root law has been derived regarding the number of bits that can be covertly and reliably transmitted over additive white Gaussian noise (AWGN) channels [3]. Specifically, with an arbitrarily low probability of being detected, one can reliably send at most $\mathcal{O}\left(\sqrt{n}\right)$ bits over $n$ uses of the channel, which further implies that the asymptotic covert rate approaches zero. The work of [3] has then been extended to different scenarios, e.g., binary symmetric channels (BSCs) [4], discrete memoryless channels (DMCs) [5], and multiple access channels (MACs) [6].

The pessimistic zero covert throughput implied by the square root law motivates some studies exploring the conditions in which positive covert throughput is achievable. It has been found that positive covert throughput can be achieved when the warden has uncertainty on the power of noise or interference at its receiver [7–10]. With a worst-case consideration from the warden's perspective, the study in [7] pointed out that the positive covert throughput is achievable when the warden has uncertainty about its receiver noise power. Also, for the scenario that the warden has uncertainty about its receiver noise power, the study in [8] showed the positive covert throughput with the consideration of the overall performance at the warden instead of the worst-case consideration from the warden's perspective. Moreover, the covert communication with a positive rate has been proved to be achievable when a friendly jammer sends artificial noise so that the warden has uncertainty on the interference at its receiver [9, 10].

In practical wireless networks, a major source of the uncertain interference or noise at the receiver is actually the ambient signals from other transmitters, and the uncertainty of the aggregate received interference at the warden may help to achieve the positive covert throughput. Thus, investigating the covert communication in wireless network scenarios is of a significant importance. Although the secrecy issue of communication in wireless network scenarios has been extensively studied in the literature, e.g., [11–13], the covertness issue of communication in wireless network scenarios has been rarely studied. The covert communication in wireless networks has been studied very recently in [10]. However, the work in [10] considered that all other communication nodes in the network are friendly helpers of the legitimate transmitter, and the closest friendly helper to the warden is selected to intentionally transmit artificial noise to confuse the warden. The requirement that all communication nodes are friendly helpers [10] is often difficult to achieve in practical wireless networks. It is worth mentioning that the covert communication in wireless networks has also been studied in [14–16]. Different from the afore-mentioned papers, [14–16] studied the covert networks through the concept of spectral outage probability, while the throughput at which the transmitter and the receiver may communicate reliably while guaranteeing a low probability of detection from the warden was not investigated. Similarly, [17] investigated the covert communication in the presence of primary networks based on the signal-to-interference-noise ratio (SINR).

In this work, we study the covert communication in wireless networks with the aid of stochastic geometry. Instead of assuming that all communication nodes in the network are friendly helpers, we consider that all nodes in the network, which are distributed according to a homogeneous Poisson point process (PPP), randomly transmit without the intention to help the covert communication, so that the aggregate received interference at the receiver is the well known shot noise. A shot noise process can represent the aggregated interferences when the nodes in the network are distributed according to a stochastic point process [18]. The investigated scenario is that a pair of communication nodes want to achieve covert communication by making use of the existing wireless network environment. The hope is to hide the communication by using the time varying interference created by all users in the large-scale wireless network. We note that the secrecy issue of wireless networks and the aggregate interference in wireless networks have been widely studied in the literature, e.g., [14, 19–29]. It is worth mentioning that our results are novel compared with those existing papers, since they have not considered the covertness issue of wireless networks.

The primary contributions of the paper are summarized as follows:

1) We comprehensively study the covert communication in wireless networks subject to shot noise [18]. We analyze the average covert probability, the connection outage probability, and the covert throughput to capture the covertness, reliability, and overall rate performances of the system, respectively.

2) We evaluate the impact of the density and the transmit power of the concurrent interferers in the network on the covert throughput. For the interference-limited network, we analytically find that the covert throughput is affected by neither the density nor the transmit power of the concurrent interferers for both cases of non-fading and fading channels.

3) While the expressions for the average covert probabilities are complicated for general values of the path loss exponent, $\alpha$, due to the stochastic geometry, we further consider the special case of $\alpha = 4$ to derive the simplified expressions for the average covert probabilities for both the non-fading and the fading cases.

4) We examine the impact of the AWGN on the performances of the covert communication in wireless networks. We find that the AWGN does not affect the average covert probability of the system in wireless networks, while it decreases the covert throughput of the system for both the non-fading and the fading cases. Moreover, we find that the covert throughput increases as the density or the transmit power of the concurrent interferers increases when the AWGN is taken into account.

Throughout the paper, we adopt the following notations: $Q(\cdot)$ denotes the tail probability of the standard normal distribution, $\Gamma(\cdot)$ denotes the standard gamma function, $\Gamma(\cdot, \cdot)$ denotes the upper incomplete gamma function, $\mathcal{CN}(\mu, \sigma^2)$ denotes the complex Gaussian distribution with mean $\mu$ and variance $\sigma^2$, $\csc(\cdot)$ denotes the cosecant function, $[x]^+$ denotes $\max\{x, 0\}$, and $\mathbb{P}$ denotes the probability measure.

## II. System Model

As shown in Figure 1, we consider a two-dimensional wireless network which consists of a transmitter, Alice, a receiver, Bob, a warden, Willie, and a Poisson field of interferers. Bob is located at a distance $d_{ab}$ from Alice and Willie is located at a distance $d_{aw}$ from Alice. We do not consider the relative motion between Alice, Bob, and Willie. The locations of concurrent interferers follow a homogeneous PPP, denoted by $\Phi$, with node density $\lambda_I$. The system is assumed to be slotted in time, and the locations of concurrent interferers remain static in a slot. We assume that the locations of current interferers change at each time slot, which happens in, e.g., mobile networks, due to either the random access or the mobility, or both. Willie would like to detect if Alice is transmitting in a slot, while Alice attempts to transmit messages to Bob with a low probability of being detected. We assume that all nodes in the network have a single antenna. We further assume that the transmitted signals from Alice and the interferers are independent zero-mean complex Gaussian signals. It is worth mentioning that the assumption of Gaussian codebooks for the interferers has negligible effect on the aggregate network interference [24]. We assume that the codeword is kept secret from Willie [3]. In addition, we assume that the pathloss effects and the fading coefficients remain constant in a slot.

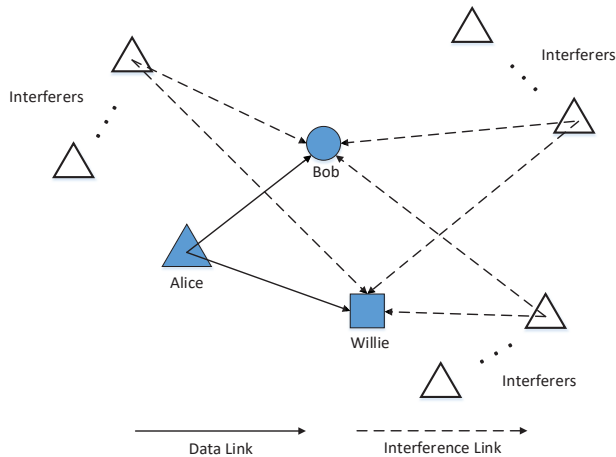

Fig. 1: Illustration of the network model for covert wireless communications with a Poisson field of interferers.

Denote the normalized transmitted signal from Alice by $x[n] \sim \mathcal{CN}(0, 1)$, where $n = 1, 2, \cdots, N$. When Alice is transmitting with power $P_a$, the received signals at Bob and Willie are, respectively, given by

$$y_b[n] = \sqrt{\frac{P_a}{d_{ab}^\alpha}} h_{ab} x[n] + v_b[n] + z_b[n] \tag{1}$$

$$y_w[n] = \sqrt{\frac{P_a}{d_{aw}^\alpha}} h_{aw} x[n] + v_w[n] + z_w[n], \tag{2}$$

where the subscripts $b$ and $w$ denote Bob and Willie, respectively, $\alpha \geq 2$ denotes the path loss exponent, $h_{ab}$ and $h_{aw}$ denote the normalized channel coefficients from Alice to the receivers, $v_b[n] \sim \mathcal{CN}(0, \sigma_{v_b}^2)$ and $v_w[n] \sim \mathcal{CN}(0, \sigma_{v_w}^2)$ denote the aggregate received interferences, and $z_b[n] \sim \mathcal{CN}(0, \sigma_{z_b}^2)$ and $z_w[n] \sim \mathcal{CN}(0, \sigma_{z_w}^2)$ denote the AWGN. We note that $v_b[n]$ and $v_w[n]$ are usually correlated in practice due to the spatial correlation of the interferences [30–33], while the correlation between $v_b[n]$ and $v_w[n]$ does not affect the analysis and results in the paper. We denote the received signal power at Bob and Willie by $P_b = P_a \left| h_{ab} \right|^2 / d_{ab}^\alpha$ and $P_w = P_a \left| h_{aw} \right|^2 / d_{aw}^\alpha$, respectively. We assume that Bob and Willie know Alice's transmit power and the instantaneous channel coefficients from Alice to themselves.

The powers of aggregate received interferences at Bob and Willie are, respectively, given by

$$\sigma_{v_b}^2 = \sum_{i \in \Phi} \frac{P_i}{d_{ib}^\alpha} \left| h_{ib} \right|^2 \tag{3}$$

$$\sigma_{v_w}^2 = \sum_{i \in \Phi} \frac{P_i}{d_{iw}^\alpha} \left| h_{iw} \right|^2, \tag{4}$$

where $P_i$ denotes the transmit power of the interferer, $d_{ib}$ and $d_{iw}$ denote the distances from the interferer to the receivers, and $h_{ib}$ and $h_{iw}$ denote the normalized channel coefficients from the interferer to the receivers. We assume that all interferers have the same transmit power [34], such that $P_i = P_I$, $\forall i \in \Phi$. Due to the random spatial distribution of the interferers and the random channel coefficients, the powers of the aggregate received interferences, $\sigma_{v_b}^2$ and $\sigma_{v_w}^2$, are random variables. With the stationary property of PPP [34], $\sigma_{v_b}^2$ and $\sigma_{v_w}^2$ in fact have the same distribution, which is denoted by $f_{\sigma_v^2}(x)$. We assume that Willie knows the distribution of the power of the aggregate received interference, $f_{\sigma_v^2}(x)$, while he does not know the power of the instantaneous aggregate received interference in a slot. Note in the slotted system model that the power of aggregate received interferences remains static for all $N$ samples in a time slot.

For the analysis in the paper, we mainly focus on the interference-limited network [34], such that $\sigma_{z_b}^2 = \sigma_{z_w}^2 = 0$. In fact, having non-zero AWGN does not affect the average covert probability of the considered system, and we will discuss the impact of having non-zero AWGN, i.e., $\sigma_{z_b}^2 \neq 0$ and $\sigma_{z_w}^2 \neq 0$, later in Section V. The average covert probability is the adopted covertness measure in this paper, which will be given later in Section II-B1.

### A. Willie's Hypothesis Test

In the interference-limited network, Willie attempts to determine whether Alice is transmitting or not by distinguishing the following two hypotheses:

$$\mathcal{H}_0 : y_w[n] = v_w[n], \tag{5a}$$

$$\mathcal{H}_1 : y_w[n] = \sqrt{\frac{P_a}{d_{aw}^\alpha}} h_{aw} x[n] + v_w[n]. \tag{5b}$$

Based on the received signal vector $\mathbf{y}_w = [y_w[1], \cdots, y_w[n]]$, Willie makes a binary decision on whether the received signal is the interference or the signal from Alice plus the interference. When Willie adopts a radiometer as the detector, the test statistic is given by $T(\mathbf{y}_w) = (1/N) \sum_{n=1}^{N} |y_w[n]|^2$. Throughout this paper, we assume that Willie adopts a radiometer as the detector [7], since the radiometer is often used in practice. A detailed justification of this assumption is given in Appendix A.

Denote $D_0$ and $D_1$ as the decision that the received signal is the interference and the decision that the received signal is the signal from Alice plus the interference, respectively. As per the mechanism of radiometer, Willie makes the decision of $D_0$ if $T(\mathbf{y}_w) \leq \gamma$, and the decision of $D_1$ if $T(\mathbf{y}_w) > \gamma$, where $\gamma > 0$ is the threshold of Willie's detector. The design of the threshold $\gamma$ is crucial for the performance of Willie's radiometer [35]. The false alarm probability and the misdetection probability are, respectively, defined as

$$P_{\text{FA}} = \mathbb{P}\left(D_1 \mid \mathcal{H}_0\right) \tag{6}$$

$$P_{\text{MD}} = \mathbb{P}\left(D_0 \mid \mathcal{H}_1\right). \tag{7}$$

Since the false alarm and the misdetection are the two types of errors for Willie's hypothesis test, following the existing studies on covert communication, the performance of Willie's hypothesis test is measured by

$$\xi = P_{\text{FA}} + P_{\text{MD}}, \tag{8}$$

where $0 \leq \xi \leq 1$. In particular, $\xi = 0$ indicates that Willie can perfectly detect the transmission without error. In contrast, $\xi = 1$ indicates that Willie cannot detect the transmission at all, which is equivalent to a random guess.

For given $\gamma$, $\sigma_{v_w}^2$, and $P_w$, we have the following equations. The likelihood functions of $T(\mathbf{y}_w)$ under $\mathcal{H}_0$ and $\mathcal{H}_1$ are, respectively, given by

$$\mathbb{P}\left(T(\mathbf{y}_w)|\mathcal{H}_0\right) = \frac{(T(\mathbf{y}_w))^{N-1}}{\Gamma(N)} \left(\frac{N}{\sigma_{v_w}^2}\right)^N \exp\left(-\frac{NT(\mathbf{y}_w)}{\sigma_{v_w}^2}\right) \tag{9}$$

and

$$\mathbb{P}\left(T(\mathbf{y}_w)|\mathcal{H}_1\right) = \frac{(T(\mathbf{y}_w))^{N-1}}{\Gamma(N)} \left(\frac{N}{P_w + \sigma_{v_w}^2}\right)^N \exp\left(-\frac{NT(\mathbf{y}_w)}{P_w + \sigma_{v_w}^2}\right). \tag{10}$$

Then, the false alarm probability and the misdetection probability for a given threshold $\gamma$ are, respectively, given by

$$P_{\text{FA}} = \frac{\Gamma\left(N, \frac{N\gamma}{\sigma_{v_w}^2}\right)}{\Gamma(N)} \tag{11}$$

and

$$P_{\text{MD}} = 1 - \frac{\Gamma\left(N, \frac{N\gamma}{P_w + \sigma_{v_w}^2}\right)}{\Gamma(N)}. \tag{12}$$

Following the widely-adopted assumption in the analysis on covert communication [7, 8, 36], we assume that $N \to \infty$ to allow Willie to get an infinite number of samples for his detection. This is an upper bound on the number of samples that Willie can get in practice. When $N \to \infty$, we have

$$\xi\left(\gamma, \sigma_{v_w}^2, P_w\right) = \begin{cases} 0, & \text{if } \sigma_{v_w}^2 \leq \gamma \leq P_w + \sigma_{v_w}^2 \\ 1, & \text{otherwise.} \end{cases} \tag{13}$$

### B. Covertness Measure

We note that, for any given detection threshold $\gamma > 0$, $\xi$ in (13) is a random variable that follows the Bernoulli distribution, due to the randomness of $\sigma_{v_w}^2$ and $P_w$. The overall performance of covertness is then measured by either the average covert probability or the covert outage probability [8]. It is worth mentioning that we do not consider the issue of signal detection at Bob, and assume that Bob knows when Alice transmits. This can be realized in practice by sharing a secret timetable between Alice and Bob.

*1) Average Covert Probability:* The average covert probability captures the average covertness of the system from a Bayesian statistics perspective [37]. The average covert probability is given by [8]

$$\bar{\xi}(P_a) = \int_0^\infty \min_{\gamma > 0} \int_0^\infty \xi\left(\gamma, \sigma_{v_w}^2, P_w\right) f_{\sigma_v^2}\left(\sigma_{v_w}^2\right) d\sigma_{v_w}^2 f_{P_w}\left(P_w\right) dP_w, \tag{14}$$

where $f_{P_w}(x)$ denotes the distribution of $P_w$. Since $P_w = P_a \left|h_{aw}\right|^2 / d_{aw}^\alpha$, $f_{P_w}(x)$ is determined by the value of $P_a/d_{aw}^\alpha$ and the distribution of $\left|h_{aw}\right|^2$. Note that $\min_\gamma$ in (14) indicates that Willie can use the optimal detection threshold $\gamma$ for his detection of Alice's transmission, since the optimal threshold at Willie minimizes his detection error, which is characterized by $\int_0^\infty \xi\left(\gamma, \sigma_{v_w}^2, P_w\right) f_{\sigma_v^2}\left(\sigma_{v_w}^2\right) d\sigma_{v_w}^2$ for given $P_w$. Hence, measuring the covertness of a system by $\bar{\xi}$ is robust from the perspective of the legitimate users, i.e., Alice and Bob.

*2) Covert Outage Probability:* The covert outage probability captures the probability that covert communication fails. From the outage-based approach, the covert outage probability is given by [8]

$$p_{\text{out}}^{\text{cv}}(P_a) = \int_0^\infty \max_{\gamma > 0} \mathbb{P}\left(\xi\left(\gamma, \sigma_{v_w}^2, P_w\right) < 1\right) f_{P_w}\left(P_w\right) dP_w. \tag{15}$$

Note that the concept of covert outage probability is similar to the concept of secrecy outage probability [38], which characterizes the probability that secure communication fails.

It is easy to find that $p_{\text{out}}^{\text{cv}} = 1 - \bar{\xi}$ in this work, due to the Bernoulli distribution of $\xi$ as $N \to \infty$. Thus, we adopt the average covert probability $\bar{\xi}$ to measure the covertness of the system. It is worth mentioning that $p_{\text{out}}^{\text{cv}} \neq 1 - \bar{\xi}$ when $\xi(\sigma_w^2, \gamma)$ starts taking values within the range of $(0, 1)$, which happens when Willie has a finite number of samples [39].

### C. Reliability Measure

Due to the fact that the power of instantaneous aggregate received interference at Bob, $\sigma_{v_b}^2$, is unknown, every possible target transmission rate is associated with an unavoidable probability of connection outage. Thus, we measure the reliability performance of the system by connection outage probability, which is given by

$$
\begin{aligned}
p_{\text{out}}^{\text{cn}}(P_a, R) &= \mathbb{P}\left(\log_2\left(1 + \frac{P_b}{\sigma_{v_b}^2}\right) < R\right) \\
&= \mathbb{P}\left(\sigma_{v_b}^2 > P_b\left(2^R - 1\right)^{-1}\right) \\
&= \int_0^\infty \int_{P_b(2^R-1)^{-1}}^\infty f_{\sigma_v^2}\left(\sigma_{v_b}^2\right) \mathrm{d}\sigma_{v_b}^2 f_{P_b}\left(P_b\right) \mathrm{d}P_b,
\end{aligned}
\tag{16}
$$

where $R$ denotes the data transmission rate, the inter integral is over the range $\sigma_{v_b}^2$ in which the connection outage happens for a given $P_b$, and the outer integral is over the range of all possible values of $P_b$.

### D. Covert Throughput

With the consideration of both covertness and reliability, the overall rate performance of the system is evaluated by covert throughput. The covert throughput captures the maximum achievable data transmission rate subject to both covertness requirement and reliability requirement, which is given by

$$
\eta = \max R, \quad \text{s.t. } \bar{\xi}(P_a) \geq 1 - \epsilon \text{ and } p_{\text{out}}^{\text{cn}}(P_a, R) \leq \delta,
\tag{17}
$$

where $1 - \epsilon$ denotes the acceptable minimum average covert probability and $\delta$ denotes the acceptable maximum connection outage probability. The feasible range of $\epsilon$ is $0 < \epsilon < 1$, since the system can satisfy any required $0 < \epsilon < 1$ by adjusting the transmit power at Alice $P_a > 0$. The feasible range of $\delta$ is $0 < \delta < 1$, since the system can satisfy any required $0 < \delta < 1$ for any given $P_a$ by adjusting the transmit rate $R > 0$. Note that the covert throughput in (17) would become the well-known outage throughput [40], if the average covert probability is not considered. If the covertness requirement is replaced by a secrecy requirement, e.g., a secrecy outage requirement, (17) can be used to measure the throughput for wireless networks with secrecy concerns.

## III. COVERT COMMUNICATION IN INTERFERENCE-LIMITED NETWORK WITHOUT FADING

In this section, we study covert communication in the interference-limited network and assume that there is no fading, such that $h_{aj} = h_{ij} = 1$, where $j = b$ or $w$ and $i \in \Phi$. Then, $P_b = P_a/d_{ab}^\alpha$ and $P_w = P_a/d_{aw}^\alpha$ are fixed for a given transmit power $P_a$. Given $|h_{iw}|^2 = 1$ and the expressions for $\sigma_{v_b}^2$ and $\sigma_{v_w}^2$ in (3) and (4), the distribution of $\sigma_{v_b}^2$ and $\sigma_{v_w}^2$ for the case of non-fading channels is found as an infinite series [41]:

$$f_{\sigma_v^2}(x) = \frac{1}{\pi x} \sum_{k=1}^{\infty} \frac{\Gamma\left(\frac{\alpha+2k}{\alpha}\right)}{k!} \left(\frac{\pi \lambda_I \Gamma\left(\frac{\alpha-2}{\alpha}\right) P_I^{\frac{2}{\alpha}}}{x^{\frac{2}{\alpha}}}\right)^k \sin\left(k\pi\left(\frac{\alpha-2}{\alpha}\right)\right), \ x > 0. \quad (18)$$

Based on (13) and (14), the average covert probability for the case of non-fading channels is given by

$$\bar{\xi} = \min_{\gamma>0} \int_0^\infty \xi\left(\gamma, \sigma_{v_w}^2\right) f_{\sigma_v^2}\left(\sigma_{v_w}^2\right) \mathrm{d}\sigma_{v_w}^2$$

$$= 1 - \max_{\gamma>0} \int_{\max\{\gamma-P_w,\ 0\}}^{\gamma} f_{\sigma_v^2}\left(\sigma_{v_w}^2\right) \mathrm{d}\sigma_{v_w}^2$$

$$\overset{(a)}{=} 1 - \max_{\gamma>P_w} \int_{\gamma-P_w}^{\gamma} f_{\sigma_v^2}\left(\sigma_{v_w}^2\right) \mathrm{d}\sigma_{v_w}^2, \quad (19)$$

where $(a)$ is due to $f_{\sigma_v^2}\left(\sigma_{v_w}^2\right) > 0$ for any $\sigma_{v_w}^2 > 0$. Substituting (18) into (19), we further derive the average covert probability for the case of non-fading channels as

$$\bar{\xi} = 1 - \frac{1}{\pi} \max_{\gamma>P_a d_{aw}^{-\alpha}} \sum_{k=1}^{\infty} \frac{\Gamma\left(\frac{2k}{\alpha}\right)\left(\pi \lambda_I \Gamma\left(\frac{\alpha-2}{\alpha}\right) P_I^{\frac{2}{\alpha}}\right)^k}{k!} \sin\left(k\pi\left(\frac{\alpha-2}{\alpha}\right)\right) \left(\frac{1}{\left(\gamma-P_a d_{aw}^{-\alpha}\right)^{\frac{2k}{\alpha}}} - \frac{1}{\gamma^{\frac{2k}{\alpha}}}\right). \quad (20)$$

Based on (16) and (18), the connection outage probability for the case of non-fading channels is given by

$$p_{\text{out}}^{\text{cn}} = \int_{P_b(2^R-1)^{-1}}^{\infty} f_{\sigma_v^2}\left(\sigma_{v_b}^2\right) \mathrm{d}\sigma_{v_b}^2 f_{P_b}\left(P_b\right)$$

$$= \frac{1}{\pi} \sum_{k=1}^{\infty} \frac{\Gamma\left(\frac{2k}{\alpha}\right)}{k!} \left(\frac{\pi \lambda_I \Gamma\left(\frac{\alpha-2}{\alpha}\right) P_I^{\frac{2}{\alpha}}}{P_a^{\frac{2}{\alpha}} d_{ab}^{-2}\left(2^R-1\right)^{-\frac{2}{\alpha}}}\right)^k \sin\left(k\pi\left(\frac{\alpha-2}{\alpha}\right)\right). \quad (21)$$

We know from the definition of connection outage probability that the connection outage probability, $p_{\text{out}}^{\text{cn}}$, monotonously increases as the transmission rate, $R$, increases or the transmit power at Alice, $P_a$, decreases. We also know from the definition of average covert probability that the average covert probability $\bar{\xi}$, decreases as $P_a$ increases. Then, the basic method to obtain the covert throughput $\eta$ is described as follows. We first solve for $P_a$ to $\bar{\xi}(P_a) = \epsilon$ to find the maximum allowable transmit power at Alice subject to the constraint of the covertness, denoted by $P_a^*$. With the derived $P_a^*$, we solve for $R$ to $p_{\text{out}}^{\text{cn}}(P_a) = \delta$ with $P_a = P_a^*$ to obtain $\eta$. However, neither the solution of $P_a$ to $\bar{\xi}(P_a) = \epsilon$ nor the solution of $R$ to $p_{\text{out}}^{\text{cn}}(P_a, R) = \delta$ is analytically tractable, since $\bar{\xi}$ and $p_{\text{out}}^{\text{cn}}$ can be given by complicated infinite series only. Thus, the derivation of the covert throughput for the case of non-fading channels is analytically intractable.

Although (20) and (21) may be complicated to calculate, nevertheless they are crucial in analyzing the impacts of the density and the transmit power of the concurrent interferers on the covert throughput in the interference-limited network with a Poisson field of interferers over non-fading channels, as will be shown in the following theorem.

*Theorem 1:* In the interference-limited network with a Poisson field of interferers over non-fading channels, the covert throughput, $\eta$, is not affected by the density of concurrent interferers, $\lambda_I$, or the transmit power at the interferers, $P_I$.

    *Proof:* See Appendix B.                                                     ∎

Note from the definition of covert throughput in (17) that Theorem 1 is valid when $P_a$ is optimally chosen to satisfy the covertness and the reliability constraints. We would like to highlight the importance of the finding in Theorem 1. In the literature on wireless networks, the density and the transmit power are usually regarded as main factors affecting the overall throughput performance; see, e.g., [34] for conventional wireless networks and [2, 13] for wireless networks with secrecy concerns. Interestingly, our analysis shows that the throughput for covert communication in the interference-limited network over non-fading channels is not affected by either the density or the transmit power of the concurrent interferers in the network. Intuitively, a large interference is beneficial for hiding a higher transmit power from Alice in Willie's uncertainty. On the other hand, a large interference leads to a higher noise for Bob. Thus, the increase of the interference power has both positive and negative effects on the overall rate performance, which is characterized by the covert throughput. Theorem 1 implies that the positive and negative effects are balanced.

## A. Special Case of $\alpha = 4$

We now consider the special case of $\alpha = 4$, which is for the case on the ground or relatively lossy environments [42]. Note that the closed-form expression for the distribution of the aggregate interference power exists for $\alpha = 4$ only [34].

When $\alpha = 4$ and $h_{ij} = 1$, $\sigma_{v_j}^2$ follows the Lévy distribution with the scalar parameter $\zeta = \pi^3 \lambda_I^2 P_I / 2$ [34], and we have

$$f_{\sigma_v^2}(x) = \frac{\pi \lambda_I \sqrt{P_I}}{2 x^{\frac{3}{2}}} \exp\left(-\frac{\pi^3 \lambda_I^2 P_I}{4x}\right), \; x > 0. \tag{22}$$

The average covert probability for the case of non-fading channels with $\alpha = 4$ is given in the following proposition.

*Proposition 1:* The average covert probability for the system over non-fading channels with $\alpha = 4$ is given by

$$\bar{\xi} = 1 - \mathrm{erf}\left(\frac{\pi^{\frac{3}{2}} \lambda_I \sqrt{P_I}}{2 \sqrt{x^o - P_a d_{aw}^{-4}}}\right) + \mathrm{erf}\left(\frac{\pi^{\frac{3}{2}} \lambda_I \sqrt{P_I}}{2 \sqrt{x^o}}\right), \tag{23}$$

where $\mathrm{erf}(\cdot)$ denotes the error function and $x^o$ is the solution of $x \geq P_w$ to

$$\left(x - P_a d_{aw}^{-4}\right)^{\frac{3}{2}} \exp\left(-\frac{\pi^3 \lambda_I^2 P_I}{4x}\right) - x^{\frac{3}{2}} \exp\left(-\frac{\pi^3 \lambda_I^2 P_I}{4\left(x - P_a d_{aw}^{-4}\right)}\right) = 0. \tag{24}$$

*Proof:* See Appendix C. ∎

The connection outage probability for the case of non-fading channels with $\alpha = 4$ is given by

$$p_{\mathrm{out}}^{\mathrm{cn}} = \int_{P_b(2^R - 1)^{-1}}^{\infty} f_{\sigma_v^2}\left(\sigma_{v_b}^2\right) \mathrm{d}\sigma_{v_b}^2 = \mathrm{erf}\left(\frac{\pi^{\frac{3}{2}} \lambda_I d_{ab}^2 \sqrt{P_I \left(2^R - 1\right)}}{2 \sqrt{P_a}}\right). \tag{25}$$

For the case of $\alpha = 4$, the solution of $P_a$ to $\bar{\xi}(P_a) = \epsilon$ can be numerically obtained and the solution of $R$ to $p_{\mathrm{out}}^{\mathrm{cn}}(P_a, R) = \delta$ is analytically tractable, which is given by

$$R = \log_2\left(1 + \frac{4 P_a \left(\mathrm{erf}^{-1}(\delta)\right)^2}{P_I \pi^3 \lambda_I^2 d_{ab}^4}\right). \tag{26}$$

With the previously discussed method and the finding in Theorem 1, we can obtain the covert throughput of the considered system over non-fading channels with $\alpha = 4$ by an iterative algorithm. For brevity, we do not present the detailed algorithm here. Instead, we summarize the basic idea of the algorithm as follows. Since the covert throughput is not affected by the density and the transmit power of concurrent interferers, we consider $\pi^3 \lambda_I^2 P_I / 4 = 1$ to simplify the expressions for (23), (24), and (26). We first adopt the one-side-search technique to decide the searching range of $P_a^*$ [43], and then adopt the bisection-search technique to find $P_a^*$. Finally, $\eta$ is obtained according to (26).

## IV. Covert Communication in Interference-Limited Network with Fading

We now study the covert communication in the case of fading channels. Denote the distribution of $|h_{ij}|^2$, where $j = b$ or $w$ and $i \in \Phi$, by $f_h(x)$. Then, the distributions of $P_b = P_a |h_{ab}|^2 / d_{ab}^\alpha$ and $P_w = P_a |h_{aw}|^2 / d_{aw}^\alpha$ for a given transmit power $P_a$ are, respectively, given by

$$f_{P_b}(x) = \frac{d_{ab}^\alpha}{P_a} f_h\left(\frac{d_{ab}^\alpha x}{P_a}\right) \tag{27}$$

$$f_{P_w}(x) = \frac{d_{aw}^\alpha}{P_a} f_h\left(\frac{d_{aw}^\alpha x}{P_a}\right). \tag{28}$$

The distribution of $\sigma_{v_b}^2$ or $\sigma_{v_w}^2$ for the case of fading channels is also found as an infinite series, which is given by [18]

$$f_{\sigma_v^2}(x) = \frac{1}{\pi x} \sum_{k=1}^{\infty} \frac{(-1)^{k+1} \Gamma\left(\frac{\alpha+2k}{\alpha}\right) \sin\left(\frac{2k\pi}{\alpha}\right)}{k!} \left(\frac{\pi \lambda_I \Gamma\left(\frac{\alpha-2}{\alpha}\right) \mathbb{E}\left\{|h_{ij}|^{\frac{4}{\alpha}}\right\} P_I^{\frac{2}{\alpha}}}{x^{\frac{2}{\alpha}}}\right)^k. \tag{29}$$

We note that $\mathbb{E}\left\{|h_{ij}|^{\frac{4}{\alpha}}\right\}$ in (29) is related to the distribution of fading channel. In the case of Rayleigh fading, we have $\mathbb{E}\left\{|h_{ij}|^{\frac{4}{\alpha}}\right\} = \Gamma\left(\frac{\alpha+2}{\alpha}\right)$, and

$$f_{\sigma_v^2}(x) = \frac{1}{\pi x} \sum_{k=1}^{\infty} \frac{(-1)^{k+1} \Gamma\left(\frac{\alpha+2k}{\alpha}\right) \sin\left(\frac{2k\pi}{\alpha}\right)}{k!} \left(\frac{2\pi^2 \lambda_I \csc\left(\frac{2\pi}{\alpha}\right) P_I^{\frac{2}{\alpha}}}{\alpha x^{\frac{2}{\alpha}}}\right)^k. \tag{30}$$

The average covert probability for the case of fading channels is given by

$$\bar{\xi} = \int_0^\infty \min_{\gamma > 0} \int_0^\infty \xi\left(\gamma, \sigma_{v_w}^2, P_w\right) f_{\sigma_v^2}\left(\sigma_{v_w}^2\right) \mathrm{d}\sigma_{v_w}^2 f_{P_w}\left(P_w\right) \mathrm{d}P_w$$

$$= \int_0^\infty \left(1 - \max_{\gamma > P_w} \int_{\gamma - P_w}^\gamma f_{\sigma_v^2}\left(\sigma_{v_w}^2\right) \mathrm{d}\sigma_{v_w}^2\right) f_{P_w}\left(P_w\right) \mathrm{d}P_w. \tag{31}$$

Substituting (28) and (29) into (31), we further derive the average covert probability for the case of fading channels as

$$\bar{\xi} = \int_0^\infty \frac{d_{aw}^\alpha}{P_a} f_h\left(\frac{d_{aw}^\alpha P_w}{P_a}\right)$$

$$\left(1 - \frac{1}{\pi} \max_{\gamma > P_w} \sum_{k=1}^{\infty} \frac{(-1)^{k+1} \Gamma\left(\frac{2k}{\alpha}\right) \sin\left(\frac{2k\pi}{\alpha}\right) \left(\pi \lambda_I \Gamma\left(\frac{\alpha-2}{\alpha}\right) \mathbb{E}\left\{|h_{ij}|^{\frac{4}{\alpha}}\right\} P_I^{\frac{2}{\alpha}}\right)^k}{k!} \left(\frac{1}{(\gamma - P_w)^{\frac{2k}{\alpha}}} - \frac{1}{\gamma^{\frac{2k}{\alpha}}}\right)\right) \mathrm{d}P_w. \tag{32}$$

Based on (16) and (29), the connection outage probability for the case of fading channels is given by

$$p_{\text{out}}^{\text{cn}} = \int_0^\infty \int_{P_b(2^R-1)^{-1}}^\infty f_{\sigma_v^2}\left(\sigma_{v_b}^2\right) \mathrm{d}\sigma_{v_b}^2 f_{P_b}\left(P_b\right) \mathrm{d}P_b,$$

$$= \int_0^\infty \frac{d_{ab}^\alpha}{P_a} f_h\left(\frac{d_{ab}^\alpha P_b}{P_a}\right) \frac{1}{\pi} \sum_{k=1}^\infty \frac{(-1)^{k+1}\Gamma\left(\frac{2k}{\alpha}\right)\sin\left(\frac{2k\pi}{\alpha}\right)}{k!} \left(\frac{\pi\lambda_I\Gamma\left(\frac{\alpha-2}{\alpha}\right)\mathbb{E}\left\{|h_{ij}|^{\frac{4}{\alpha}}\right\}P_I^{\frac{2}{\alpha}}}{\left(P_b\left(2^R-1\right)^{-1}\right)^{\frac{2}{\alpha}}}\right)^k \mathrm{d}P_b. \tag{33}$$

The basic method to obtain the covert throughput $\eta$ for the case of fading channels is the same as the non-fading channels. We first solve for $P_a$ to $\bar{\xi}(P_a) = \epsilon$ to find the maximum allowable transmit power at Alice subject to the constraint on the covertness, denoted by $P_a^*$. With the derived $P_a^*$, we then solve for $R$ to $p_{\text{out}}^{\text{cn}}(P_a) = \delta$ with $P_a = P_a^*$ to obtain $\eta$. Since neither the solution of $P_a$ to $\bar{\xi}(P_a) = \epsilon$ nor the solution of $R$ to $p_{\text{out}}^{\text{cn}}(P_a, R) = \delta$ is analytically tractable, the derivation of the covert throughput for the case of fading channels is analytically intractable.

Similar to the analysis in Section III, although (32) and (33) may be complicated to calculate, nevertheless they are crucial in analyzing the impacts of the density and the transmit power of the concurrent interferers on the covert throughput in the interference-limited network with a Poisson field of interferers over fading channels, as will be shown in the following theorem.

*Theorem 2:* In the interference-limited network with a Poisson field of interferers over fading channels, the covert throughput, $\eta$, is not affected by the density of concurrent interferers, $\lambda_I$, or the transmit power at the interferers, $P_I$.

*Proof:* See Appendix D. ∎

Again, we highlight the importance of the finding in Theorem 2. Different from the conventional impression that the density and the transmit power are main factors affecting the overall throughput performance in wireless networks [2, 13, 34], our analysis finds that the throughput for covert communication in the interference-limited network over fading channels is affected by neither the density nor the transmit power of the concurrent interferers in the network.

### A. Rayleigh Fading with $\alpha = 4$

We now consider the case of Rayleigh fading, such that $h_{ij} \sim \mathcal{CN}(0, 1)$, where $j = b$ or $w$ and $i \in \Phi$. Again, we consider the special case of $\alpha = 4$, since the closed-form expression for the distribution of the aggregate interference power exists for $\alpha = 4$ only [34].

The distributions of $P_b$ and $P_w$ for the case of Rayleigh fading channels with $\alpha = 4$ are, respectively, given by

$$f_{P_b}(x) = \frac{d_{ab}^4}{P_a} \exp\left(-\frac{d_{ab}^4 x}{P_a}\right), \ x > 0 \tag{34}$$

and

$$f_{P_w}(x) = \frac{d_{aw}^4}{P_a} \exp\left(-\frac{d_{aw}^4 x}{P_a}\right), \ x > 0. \tag{35}$$

When $\alpha = 4$ and $h_{ij} \sim \mathcal{CN}(0,1)$, $\sigma_{v_j}^2$ follows the Lévy distribution with the scalar parameter $\zeta = \pi^4 \lambda_I^2 P_I / 8$ [34], and we have

$$f_{\sigma_v^2}(x) = \frac{\pi^{\frac{3}{2}} \lambda_I \sqrt{P_I}}{4 x^{\frac{3}{2}}} \exp\left(-\frac{\pi^4 \lambda_I^2 P_I}{16 x}\right), \ x > 0. \tag{36}$$

The average covert probability for the case of Rayleigh fading channels with $\alpha = 4$ is given in the following proposition.

*Proposition 2:* The average covert probability for the system over Rayleigh fading channels with $\alpha = 4$ is given by

$$\bar{\xi} = 1 - \frac{d_{aw}^4}{P_a} \int_0^\infty \left( \text{erf}\left(\frac{\pi^2 \lambda_I \sqrt{P_I}}{4\sqrt{x^o - P_w}}\right) - \text{erf}\left(\frac{\pi^2 \lambda_I \sqrt{P_I}}{4\sqrt{x^o}}\right) \right) \exp\left(-\frac{P_w d_{aw}^4}{P_a}\right) \mathrm{d}P_w, \tag{37}$$

where $x^o$ is the solution of $x \geq P_w$ to

$$(x - P_w)^{\frac{3}{2}} \exp\left(-\frac{\pi^4 \lambda_I^2 P_I}{16x}\right) - x^{\frac{3}{2}} \exp\left(-\frac{\pi^4 \lambda_I^2 P_I}{16(x - P_w)}\right) = 0. \tag{38}$$

*Proof:* See Appendix E. ∎

The connection outage probability for the case of Rayleigh fading channels with $\alpha = 4$ is given by

$$p_{\text{out}}^{\text{cn}} = \int_0^\infty \int_{P_b(2^R - 1)^{-1}}^\infty f_{\sigma_v^2}\left(\sigma_{v_b}^2\right) \mathrm{d}\sigma_{v_b}^2 f_{P_b}(P_b) \mathrm{d}P_b$$
$$= 1 - \exp\left(-\frac{\pi^2 \lambda_I d_{ab}^2 \sqrt{P_I(2^R - 1)}}{2\sqrt{P_a}}\right). \tag{39}$$

Similar to the case of non-fading channels with $\alpha = 4$, the solution of $P_a$ to $\bar{\xi}(P_a) = \epsilon$ can be numerically obtained and the solution of $R$ to $p_{\text{out}}^{\text{cn}}(P_a, R) = \delta$ is analytically tractable, which is given by

$$R = \log_2\left(1 + \frac{4 P_a \left(\ln(1 - \delta)\right)^2}{P_I \pi^4 \lambda_I^2 d_{ab}^4}\right). \tag{40}$$

The covert throughput of the considered system over Rayleigh fading channels with $\alpha = 4$ can be obtained from an algorithm similar to the algorithm discussed in Section III-A, and the detailed algorithm is omitted here to avoid repetition.

## V. Covert Communication in Interference Network with AWGN

In the previous sections, we focus on the scenario of interference-limited network, such that the AWGN at the receivers are negligible, i.e., $\sigma_{z_b}^2 = \sigma_{z_w}^2 = 0$. In this section, we take the non-zero AWGN into account. We assume that $\sigma_{z_b}^2$ and $\sigma_{z_w}^2$ are known at Bob and Willie, respectively.

### A. Impact of AWGN on System Performance

Subject to the AWGN, Willie attempts to determine whether Alice is transmitting or not by distinguishing the following two hypotheses now:

$$\mathcal{H}_0 : y_w[n] = v_w[n] + z_w[n], \tag{41a}$$

$$\mathcal{H}_1 : y_w[n] = \sqrt{\frac{P_a}{d_{ai}^\alpha}} h_{ai} x[n] + v_w[n] + z_w[n]. \tag{41b}$$

Based on the received vector $\mathbf{y}_w = [y_w[1], \cdots, y_w[n]]$, Willie makes a binary decision on whether the received signal is the interference plus the noise or the signal from Alice plus the interference and the noise. With the radiometer as the detector, the test statistic is $T(\mathbf{y}_w) = (1/N) \sum_{n=1}^N |y_w[n]|^2$. The false alarm probability and the misdetection probability for a given threshold $\gamma$ are now, respectively, given by

$$P_{\text{FA}} = \frac{\Gamma\left(N, \frac{N\gamma}{\sigma_{v_w}^2 + \sigma_{z_w}^2}\right)}{\Gamma(N)} \tag{42}$$

and

$$P_{\text{MD}} = 1 - \frac{\Gamma\left(N, \frac{N\gamma}{P_w + \sigma_{v_w}^2 + \sigma_{z_w}^2}\right)}{\Gamma(N)}. \tag{43}$$

As $N \to \infty$, the performance of Willie's hypothesis test is given by

$$\xi\left(\sigma_{v_w}^2, \gamma, P_w, \sigma_{z_w}^2\right) = \begin{cases} 0, & \text{if } \sigma_{v_w}^2 + \sigma_{z_w}^2 \leq \gamma \leq P_w + \sigma_{v_w}^2 + \sigma_{z_w}^2 \\ 1, & \text{otherwise,} \end{cases} \tag{44}$$

and the average covert probability is given by

$$\bar{\xi}(P_a, \sigma_{z_w}^2) = \int_0^\infty \min_{\gamma > 0} \int_0^\infty \xi\left(\gamma, \sigma_{v_w}^2, P_w, \sigma_{z_w}^2\right) f_{\sigma_v^2}\left(\sigma_{v_w}^2\right) d\sigma_{v_w}^2 f_{P_w}\left(P_w\right) dP_w. \tag{45}$$

From (44) and (45), we have the following proposition on the impact of AWGN on the average covert probability of the considered system.

*Proposition 3:* Having the non-zero AWGN at Willie does not affect the average covert probability, i.e., $\bar{\xi}(P_a, \sigma_{z_w}^2 = a) = \bar{\xi}(P_a, \sigma_{z_w}^2 = 0)$ for any $a > 0$.

*Proof:* Note that Willie knows the noise variance, $\sigma_{z_w}^2$. Thus, for any given $\sigma_{z_w}^2$, Willie can set his optimal power detection threshold as the optimal threshold for the case of no AWGN plus the noise variance $\sigma_{z_w}^2$. Then, the resulted average covert probability for the case of non-zero AWGN remains the same as that for the case of no AWGN. ∎

Based on Proposition 3, we note that the expressions for the average covert probabilities are not affected by the non-zero AWGN. Hence, the average covert probabilities for the case of AWGN channels and the case of fading channels with AWGN are still given by (20) and (32), respectively.

With the non-zero AWGN at Bob, the connection outage probability is given by

$$
\begin{aligned}
p_{\text{out}}^{\text{cn}}(P_a, R, \sigma_{z_b}^2) &= \mathbb{P}\left( \log_2\left( 1 + \frac{P_b}{\sigma_{v_b}^2 + \sigma_b^2} \right) < R \right) \\
&\overset{(a)}{=} \mathbb{P}\left( \sigma_{v_b}^2 > \left[ P_b \left( 2^R - 1 \right)^{-1} - \sigma_{z_b}^2 \right]^+ \right) \\
&= \int_0^\infty \int_{\left[ P_b(2^R-1)^{-1} - \sigma_{z_b}^2 \right]^+}^\infty f_{\sigma_v^2}\left( \sigma_{v_b}^2 \right) \, \mathrm{d}\sigma_{v_b}^2 f_{P_b}\left( P_b \right) \, \mathrm{d}P_b,
\end{aligned}
\tag{46}
$$

where $(a)$ is due to the fact that $\sigma_{v_b}^2$ is nonnegative. From (46), we find that having non-zero AWGN at Bob increases the connection outage probability, since $p_{\text{out}}^{\text{cn}}(P_a, R, \sigma_{z_b}^2)$ increases as $\sigma_{z_b}^2$ increases. Following the similar analysis given in Sections III and IV, we obtain the connection outage probabilities for the case of AWGN channels and the case of fading channels with AWGN, respectively, as

$$
p_{\text{out}}^{\text{cn}} = \frac{1}{\pi} \sum_{k=1}^\infty \frac{\Gamma\left( \frac{2k}{\alpha} \right)}{k!} \left( \frac{\pi \lambda_I \Gamma\left( \frac{\alpha-2}{\alpha} \right) P_I^{\frac{2}{\alpha}}}{\left( \left[ P_a d_{ab}^{-\alpha} \left( 2^R - 1 \right)^{-1} - \sigma_{z_b}^2 \right]^+ \right)^{\frac{2}{\alpha}}} \right)^k \sin\left( k\pi \left( \frac{\alpha-2}{\alpha} \right) \right)
\tag{47}
$$

and

$$
p_{\text{out}}^{\text{cn}} = \int_0^\infty \frac{d_{ab}^\alpha}{P_a} f_h\left( \frac{d_{ab}^\alpha P_b}{P_a} \right) \frac{1}{\pi} \sum_{k=1}^\infty \frac{(-1)^{k+1} \Gamma\left( \frac{2k}{\alpha} \right) \sin\left( \frac{2k\pi}{\alpha} \right)}{k!} \left( \frac{\pi \lambda_I \Gamma\left( \frac{\alpha-2}{\alpha} \right) \mathbb{E}\left\{ |h_{ij}|^{\frac{4}{\alpha}} \right\} P_I^{\frac{2}{\alpha}}}{\left( \left[ P_b \left( 2^R - 1 \right)^{-1} - \sigma_{z_b}^2 \right]^+ \right)^{\frac{2}{\alpha}}} \right)^k \mathrm{d}P_b.
\tag{48}
$$

Recall that the covert throughput is given by

$$
\eta = \max R, \quad \text{s.t. } \bar{\xi}(P_a) \geq 1 - \epsilon \text{ and } p_{\text{out}}^{\text{cn}}(P_a, R, \sigma_{z_b}^2) \leq \delta.
\tag{49}
$$

Based on the analysis above, we note that having the non-zero AWGN does not affect the covertness constraint $\bar{\xi}(P_a) \geq 1 - \epsilon$, while we have to decrease the transmission rate $R$ to satisfy the reliability constraint $p_{\text{out}}^{\text{cn}}(P_a, R, \sigma_{z_b}^2) \leq \delta$ as the non-zero AWGN is taken into consideration. Thus, having the non-zero AWGN decreases the achievable covert throughput compared with the interference-limited network.

## B. Impact of Network Parameters with AWGN Consideration

In Sections III and IV, we have obtained the important findings on the impact of network parameters on the covert throughput in the interference-limited networks in Theorems 1 and 2. Based on Theorems 1 and 2 and the analysis in Section V-A, we have the following corollaries regarding the impact of network parameters on the covert throughput in the wireless network with the consideration of AWGN.

*Corollary 1:* In the network with a Poisson field of interferers over AWGN channels, the covert throughput, $\eta$, increases as the density of concurrent interferers, $\lambda_I$, or the transmit power at interferers, $P_I$, increases.

*Proof:* See Appendix F. ∎

*Corollary 2:* In the network with a Poisson field of interferers over fading channels with the AWGN, the covert throughput, $\eta$, increases as the density of concurrent interferers, $\lambda_I$, or the transmit power at interferers, $P_I$, increases.

*Proof:* The proof of Corollary 2 is similar to that of Corollary 1, i.e., Appendix F. The details are omitted here for brevity. ∎

We highlight that important insights can be drawn from the results in Corollaries 1 and 2 and Theorems 1 and 2 together. That is, while increasing the interference helps to improve covert throughput for the system with AWGN, such a benefit is diminishing as the interference increases and the system becomes interference-limited.

## VI. NUMERICAL RESULTS

In this section, we present numerical results. Unless otherwise stated, we set the transmit power at interferers as $P_I = 20$ dBm, the density of concurrent interferers as $\lambda_I = 10^{-3}$, the distance between Alice and Bob as $d_{ab} = 2$, the distance between Alice and Warden as $d_{aw} = 5$, the covertness requirement as $\epsilon = 0.1$, the reliability requirement as $\delta = 0.1$, the path loss exponent as $\alpha = 4$, and the power of AWGN as $\sigma_{z_b}^2 = \sigma_{z_w}^2 = \sigma_z^2 = -50$ dBm. The results without and with fading are both shown. For the result with fading, we consider the Rayleigh fading.

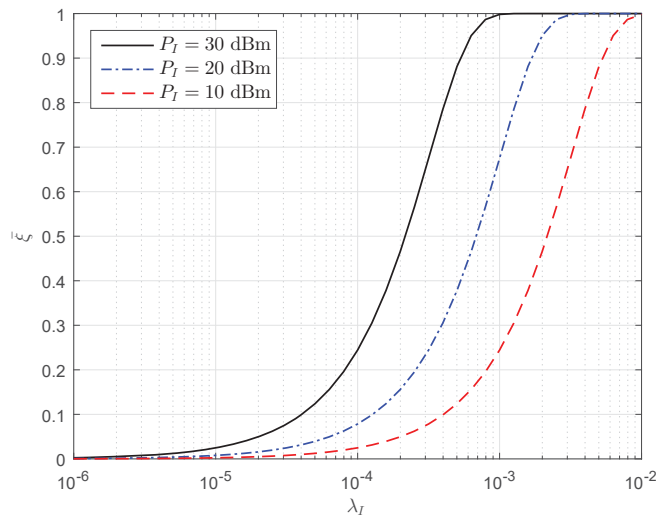

Fig. 2: Without fading: average covert probability versus density of concurrent interferers.

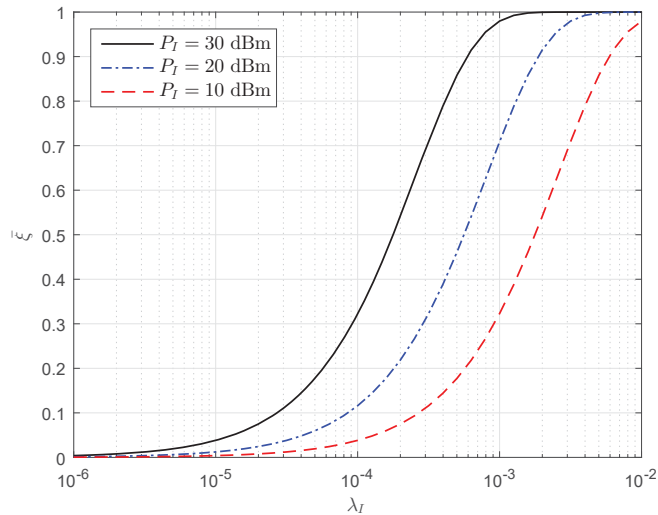

Fig. 3: With fading: average covert probability versus density of concurrent interferers.

We first demonstrate the average covert probability, $\bar{\xi}$, for networks with different densities of concurrent interferers, $\lambda_I$, and different transmit powers at interferers, $P_I$. Figures 2 and 3 plot $\bar{\xi}$ versus $\lambda_I$ with different values of $P_I$ for the case without fading and the case with fading, respectively. The transmit power at Alice is set as $P_a = 0$ dBm. As shown in both figures, $\bar{\xi}$ increases until it approaches 1, as $\lambda_I$ or $P_I$ increases, which indicates that the average covert probability improves as the interference grows. In addition, we intuitively know that the reliability performance worsens for larger interference. Mathematically, the connection outage probability, $p_{\text{out}}^{\text{cn}}$, increases as $\lambda_I$ or $P_I$ increases. Thus, the increase of interference has a positive effect on the average covert probability and a negative effect on the reliability performance, and it is not easy to determine from intuitions the impacts of $\lambda_I$ and $P_I$ on the overall rate performance of the system, i.e, the covert throughput, $\eta$.

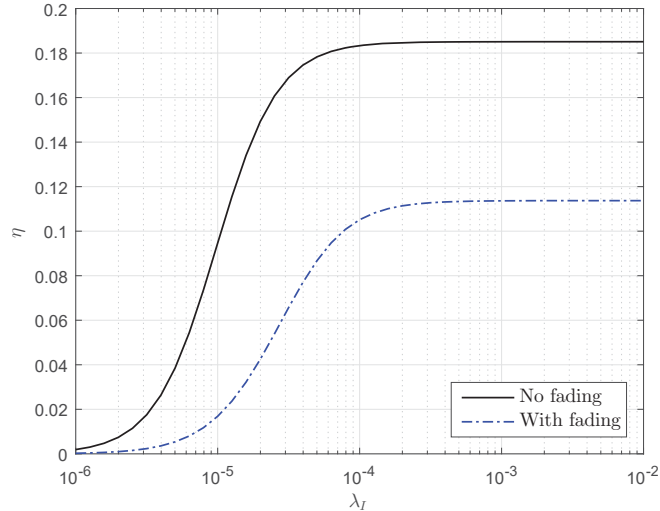

Fig. 4: Covert throughput versus density of concurrent interferers.

We now show the impact of the density of concurrent interferers, $\lambda_I$, on the covert throughput, $\eta$. Figure 4 plots $\eta$ versus $\lambda_I$ for cases with and without the fading. As depicted in the figure, when $\lambda$ is relatively small, $\eta$ increases as $\lambda_I$ increases. That observation is consistent with the findings in Corollaries 1 and 2. That is, with the consideration AWGN, the covert throughput increases as the density of concurrent interferers increases. We also note from the figure that, when $\lambda$ becomes relatively large, $\eta$ remains (almost) constant as $\lambda_I$ further increases. That observation is consistent with the findings in Theorems 1 and 2. As the density of concurrent interferers becomes relatively large, the network becomes interference-limited, and the covert throughput is not affected by the density of concurrent interferers in the interference-limited network.

We then present the impact of the transmit power at interferers, $P_I$, on the covert throughput, $\eta$. Figure 5 plots $\eta$ versus $P_I$ for cases with and without the fading. Similar to the observations in Figure 4, we note that $\eta$ increases as $\lambda_I$ increases, when $P_I$ is relatively small. When $P_I$ becomes relatively large, $\eta$ remains (almost) constant as $P_I$ further increases. Those observations are also consistent with the findings in Theorems 1 and 2 and Corollaries 1 and 2.

We further illustrate the covert throughput, $\eta$, subject to different levels of covertness requirement, $\epsilon$, where the covertness constraint is $\bar{\xi} \geq 1 - \epsilon$. Figure 6 plots $\eta$ versus $\epsilon$ for cases with and without the fading. As the figure shows, $\eta$ increases as $\epsilon$ increases, which indicates that a larger covert throughput can be achieved as the requirement on covertness becomes looser.

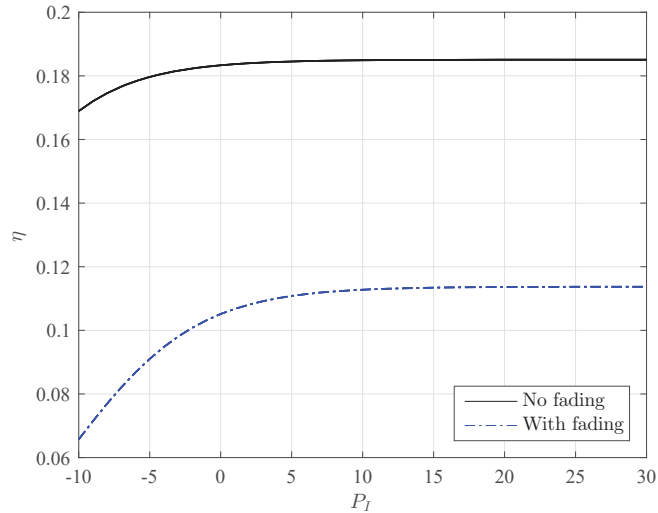

Fig. 5: Covert throughput versus transmit power at interferers.

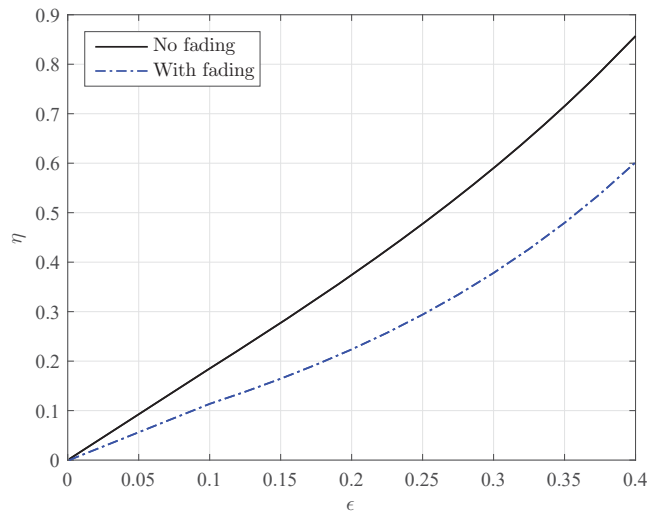

Fig. 6: Covert throughput versus covertness requirement.

Finally, we show the effect of having AWGN at the receivers on the covert throughput. Figure 7 plots the covert throughput, $\eta$, versus the power of AWGN, $\sigma_z^2$. We note from the figure that $\eta$ remains (almost) constant as $\sigma_z^2$ increases, when $\sigma_z^2$ is relatively small. This is because the network is interference-limited when $\sigma_z^2$ is small, and the small increase of $\sigma_z^2$ has little impact on the system in the interference-limited network. We further note that $\eta$ decreases as $\sigma_z^2$ further increases, which is consistent with the analysis given in Section V-A.

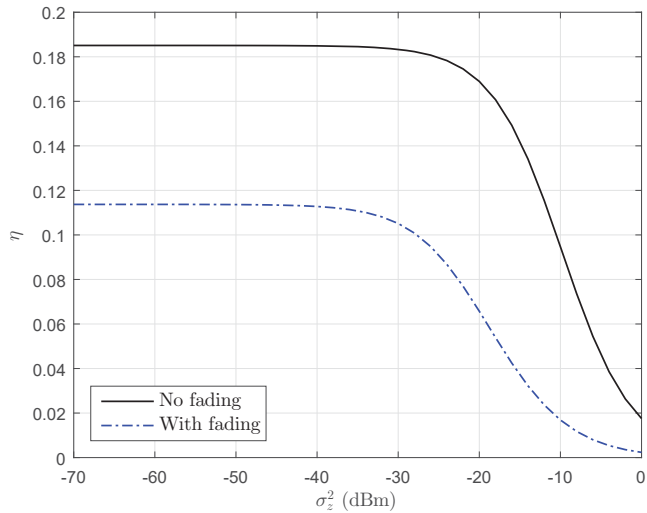

Fig. 7: Covert throughput versus power of AWGN.

## VII. Conclusion and Future Work

In this work, we have studied the covert communication in wireless networks with a Poisson field of interferers. We have analyzed the average covert probability, the connection outage probability, and the covert throughput of the system for the non-fading and the fading cases. Although the expressions for the average covert probability and the connection outage probability with general values of the path loss exponent may be complicated to calculate due to the stochastic geometry, nevertheless they are crucial in analyzing the impacts of the density and the transmit power of the concurrent interferers on the covert throughput. In particular, we have analytically found that the covert throughput is affected by neither the density nor the transmit power of the concurrent interferers in interference-limited networks for both cases of non-fading and fading channels. We have further derived the simplified expressions for the average covert probability and the connection outage probability by considering $\alpha = 4$ for both cases. Moreover, we have found that, when the AWGN is not negligible, the covert throughput increases as the density or the transmit power of the concurrent interferers increases.

While this paper has focused on the scenario where the covertness constraint is imposed on a single link, a natural extension is to investigate the scenario where the covertness constraint is imposed on all communication links in the wireless network.

## Appendix A

### Justification of assuming that Willie adopts a radiometer

In order to justify the assumption that Willie adopts a radiometer as the detector, in the following, we show that $T(\mathbf{y}_w)$ is a sufficient statistic for Willie's hypothesis test.

We first rewrite the received signal at Willie as

$$y_w[n] = Ax[n]\theta + v_w[n] + z_w[n], \tag{50}$$

where $A = \sqrt{\frac{P_A}{d_{aw}^\alpha}}h_{aw}$ and $\theta$ represents the indictor of whether Alice is transmitting or not. If Alice transmits in the slot, $\theta = 1$. Otherwise, $\theta = 0$. Willie's hypothesis test is in fact to determine if $\theta = 1$ or $0$ based on his observation $\mathbf{y}_w = [y_w[1], \cdots, y_w[n]]$. For a given slot, we note that $x[n] \sim \mathcal{CN}(0,1)$, $v_w[n] \sim \mathcal{CN}\left(0, \sigma_{v_w}^2\right)$, $z_w[n] \sim \mathcal{CN}\left(0, \sigma_{z_w}^2\right)$, $A = \sqrt{\frac{P_A}{d_{aw}^\alpha}}h_{aw}$ remains constant, and $\theta$ remains constant. Thus for a given slot, $y_w[n]$ is a complex Gaussian random variable, and $y_w[n] \sim \mathcal{CN}\left(0, \sigma_\theta^2\right)$, where $\sigma_\theta^2 = |A|^2\theta + \sigma_{v_w}^2 + \sigma_{z_w}^2$.

With a radiometer, Willie's test statistic is $T(\mathbf{y}_w) = (1/N)\sum_{n=1}^N |y_w[n]|^2$. According to Fisher-Neyman factorization theorem, the test statistic $T(\mathbf{y}_w) = (1/N)\sum_{n=1}^N |y_w[n]|^2$ is sufficient for underlying parameters $\theta$ if and only if nonnegative functions $f$ and $g$ can be found such that

$$\mathbb{P}\left(\mathbf{y}_w \mid \theta\right) = f(\mathbf{y}_w)g_\theta\left(T(\mathbf{y}_w)\right), \tag{51}$$

where $f(\mathbf{y}_w)$ does not depend on $\theta$ and $g_\theta\left(T(\mathbf{y}_w)\right)$, which does depend on $\theta$, depends on $\mathbf{y}_w$ only through $T(\mathbf{y}_w)$. Thus, in what follows, we show that the functions $f$ and $g$ above can be found. We have

$$\mathbb{P}\left(\mathbf{y}_w \mid \theta\right) = \int \prod_{n=1}^N \frac{1}{\sqrt{2\pi\sigma_\theta^2}}\exp\left(-\frac{|y_w[n]|^2}{2\sigma_\theta^2}\right)f_{\sigma_{v_w}^2}\left(\sigma_{v_w}^2\right)\mathrm{d}\sigma_{v_w}^2$$

$$= \int \left(2\pi\sigma_\theta^2\right)^{-\frac{N}{2}}\exp\left(-\frac{N}{2\sigma_\theta^2}T(\mathbf{y}_w)\right)f_{\sigma_{v_w}^2}\left(\sigma_{v_w}^2\right)\mathrm{d}\sigma_{v_w}^2. \tag{52}$$

We can find that $f(\mathbf{y}_w) = 1$ and $g_\theta\left(T(\mathbf{y}_w)\right) = \int \left(2\pi\sigma_\theta^2\right)^{-\frac{N}{2}}\exp\left(-\frac{N}{2\sigma_\theta^2}T(\mathbf{y}_w)\right)f_{\sigma_{v_w}^2}\left(\sigma_{v_w}^2\right)\mathrm{d}\sigma_{v_w}^2$. Note that $g_\theta\left(T(\mathbf{y}_w)\right) = \int \left(2\pi\sigma_\theta^2\right)^{-\frac{N}{2}}\exp\left(-\frac{N}{2\sigma_\theta^2}T(\mathbf{y}_w)\right)f_{\sigma_{v_w}^2}\left(\sigma_{v_w}^2\right)\mathrm{d}\sigma_{v_w}^2$ depends on $\mathbf{y}_w$ only through $T(\mathbf{y}_w)$, no matter what the explicit expression for $f_{\sigma_{v_w}^2}(x)$ is. This shows that $T(\mathbf{y}_w)$ is a sufficient statistic for Willie's hypothesis test.

APPENDIX B

PROOF OF THEOREM 1

We first show the proof that $\eta$ is not affected by $\lambda_I$. With $\bar{\xi} = 1 - \epsilon$ and $p_{\text{out}}^{\text{cn}} = \delta$, the covert throughput is given as the solution of $R$ to

$$\delta = \sum_{k=1}^{\infty} f_1(k) \left( \frac{\lambda_I}{(P_a^*)^{\frac{2}{\alpha}} d_{ab}^{-2} (2^R - 1)^{-\frac{2}{\alpha}}} \right)^k, \tag{53}$$

where

$$f_1(k) = \frac{\Gamma\left(\frac{2k}{\alpha}\right) \left(\pi \Gamma\left(\frac{\alpha-2}{\alpha}\right) P_I^{\frac{2}{\alpha}}\right)^k}{k! \pi} \sin\left(k\pi \left(\frac{\alpha-2}{\alpha}\right)\right), \tag{54}$$

$P_a^*$ is the solution of $P_a$ to

$$\epsilon = \max_{\gamma > P_a d_{aw}^{-\alpha}} \sum_{k=1}^{\infty} f_1(k) \left( \left( \frac{\lambda_I}{(\gamma - P_a d_{aw}^{-\alpha})^{\frac{2}{\alpha}}} \right)^k - \left( \frac{\lambda_I}{(\gamma)^{\frac{2}{\alpha}}} \right)^k \right). \tag{55}$$

We use $P_{a1}^*$ and $\eta_1$ to represent the solution of $P_a$ to (55) with $\lambda_I = \lambda_{I1} > 0$ and the covert throughput for $\lambda_I = \lambda_{I1} > 0$, i.e., the solution of $R$ to (53) with $\lambda_I = \lambda_{I1} > 0$ and $P_a^* = P_{a1}^*$, respectively. For any real positive value $u$, we then have

$$\begin{aligned}
\epsilon &= \max_{\gamma > P_{a1}^* d_{aw}^{-\alpha}} \sum_{k=1}^{\infty} f_1(k) \left( \left( \frac{\lambda_{I1}}{(\gamma - P_{a1}^* d_{aw}^{-\alpha})^{\frac{2}{\alpha}}} \right)^k - \left( \frac{\lambda_{I1}}{\gamma^{\frac{2}{\alpha}}} \right)^k \right) \\
&= \max_{\gamma > P_{a1}^* d_{aw}^{-\alpha}} \sum_{k=1}^{\infty} f_1(k) \left( \left( \frac{u\lambda_{I1}}{(u^{\frac{\alpha}{2}}\gamma - u^{\frac{\alpha}{2}} P_{a1}^* d_{aw}^{-\alpha})^{\frac{2}{\alpha}}} \right)^k - \left( \frac{u\lambda_{I1}}{(u^{\frac{\alpha}{2}}\gamma)^{\frac{2}{\alpha}}} \right)^k \right) \\
&\stackrel{(a)}{=} \max_{\gamma > u^{\frac{\alpha}{2}} P_{a1}^* d_{aw}^{-\alpha}} \sum_{k=1}^{\infty} f_1(k) \left( \left( \frac{u\lambda_{I1}}{(\gamma - u^{\frac{\alpha}{2}} P_{a1}^* d_{aw}^{-\alpha})^{\frac{2}{\alpha}}} \right)^k - \left( \frac{u\lambda_{I1}}{\gamma^{\frac{2}{\alpha}}} \right)^k \right),
\end{aligned} \tag{56}$$

and

$$\begin{aligned}
\delta &= \sum_{k=1}^{\infty} f_1(k) \left( \frac{\lambda_{I1}}{(P_{a1}^*)^{\frac{2}{\alpha}} d_{ab}^{-2} (2^{\eta_1} - 1)^{-\frac{2}{\alpha}}} \right)^k \\
&= \sum_{k=1}^{\infty} f_1(k) \left( \frac{u\lambda_{I1}}{(u^{\frac{\alpha}{2}} P_{a1}^*)^{\frac{2}{\alpha}} d_{ab}^{-2} (2^{\eta_1} - 1)^{-\frac{2}{\alpha}}} \right)^k,
\end{aligned} \tag{57}$$

where $(a)$ in (56) is derived by interchanging $u^{\frac{\alpha}{2}}\gamma$ to $\gamma$.

Now, define $\lambda_{I2} = u\lambda_{I1}$. We use $P_{a2}^*$ and $\eta_2$ to represent the solution of $P_a$ to (55) with $\lambda_I = \lambda_{I2}$ and the covert throughput for $\lambda_I = \lambda_{I2}$, i.e., the solution of $R$ to (53) with $\lambda_I = \lambda_{I2}$ and $P_a^* = P_{a2}^*$, respectively. We then have

$$
\begin{aligned}
\epsilon &= \max_{\gamma > P_{a2}^* d_{aw}^{-\alpha}} \sum_{k=1}^{\infty} f_1(k) \left( \left( \frac{\lambda_{I2}}{(\gamma - P_{a2}^* d_{aw}^{-\alpha})^{\frac{2}{\alpha}}} \right)^k - \left( \frac{\lambda_{I2}}{\gamma^{\frac{2}{\alpha}}} \right)^k \right) \\
&= \max_{\gamma > P_{a2}^* d_{aw}^{-\alpha}} \sum_{k=1}^{\infty} f_1(k) \left( \left( \frac{u\lambda_{I1}}{(\gamma - P_{a2}^* d_{aw}^{-\alpha})^{\frac{2}{\alpha}}} \right)^k - \left( \frac{u\lambda_{I1}}{\gamma^{\frac{2}{\alpha}}} \right)^k \right).
\end{aligned}
\tag{58}
$$

Comparing (56) and (58), we note that $P_{a2}^* = u^{\frac{\alpha}{2}} P_{a1}^*$. With $P_{a2}^* = u^{\frac{\alpha}{2}} P_{a1}^*$, we further have

$$
\begin{aligned}
\delta &= \sum_{k=1}^{\infty} f_1(k) \left( \frac{\lambda_{I2}}{(P_{a2}^*)^{\frac{2}{\alpha}} d_{ab}^{-2} (2^{\eta_2} - 1)^{-\frac{2}{\alpha}}} \right)^k \\
&= \sum_{k=1}^{\infty} f_1(k) \left( \frac{u\lambda_{I1}}{\left( u^{\frac{\alpha}{2}} P_{a1}^* \right)^{\frac{2}{\alpha}} d_{ab}^{-2} (2^{\eta_2} - 1)^{-\frac{2}{\alpha}}} \right)^k,
\end{aligned}
\tag{59}
$$

Comparing (57) and (59), we note that $\eta_2 = \eta_1$. Thus, the system with $\lambda_I = \lambda_{I1}$ and the system with $\lambda_I = \lambda_{I2} = u\lambda_{I1}$ have the same covert throughput. In other words, the covert throughput, $\eta$, remains the same for systems with different values of $\lambda_I$. The proof that $\eta$ is not affected by $P_I$ can be shown with the similar method as given above, and hence is omitted here. This completes the proof of Theorem 1.

## APPENDIX C

### PROOF OF PROPOSITION 1

Substituting into (22) into (19), we have

$$
\bar{\xi} = 1 - \max_{\gamma > P_w} \int_{\gamma - P_w}^{\gamma} \frac{\pi \lambda \sqrt{P_I}}{2(\sigma_{v_w}^2)^{\frac{3}{2}}} \exp \left( -\frac{\pi^3 \lambda^2 P_I}{4\sigma_{v_w}^2} \right) \mathrm{d}\sigma_{v_w}^2.
\tag{60}
$$

To derive the average covert probability, we first need to determine the optimal threshold of Willie's detector, which is given by

$$
\gamma^o = \arg\max_{\gamma > P_w} T_1(\gamma),
\tag{61}
$$

where $T_1(\gamma) = \int_{\gamma - P_w}^{\gamma} \frac{\pi \lambda \sqrt{P_I}}{2(\sigma_{v_w}^2)^{\frac{3}{2}}} \exp\left(-\frac{\pi^3 \lambda^2 P_I}{4\sigma_{v_w}^2}\right) \mathrm{d}\sigma_{v_w}^2 = \mathrm{erf}\left(\frac{\lambda_I \sqrt{\pi^3 P_I}}{2\sqrt{\gamma - P_w}}\right) - \mathrm{erf}\left(\frac{\lambda_I \sqrt{\pi^3 P_i}}{2\sqrt{\gamma}}\right)$. The average covert probability can be then rewrite as

$$
\bar{\xi} = 1 - T_1(\gamma^o).
\tag{62}
$$

Taking the first derivative of $T(\gamma)$ with respect to $\gamma$, we have

$$T_1'(\gamma) = \frac{\sqrt{B_1}}{\sqrt{\pi}} \left( \frac{\exp(-\frac{B_1}{\gamma})}{\gamma^{\frac{3}{2}}} - \frac{\exp(-\frac{B_1}{\gamma - P_w})}{(\gamma - P_w)^{\frac{3}{2}}} \right), \tag{63}$$

where $B_1 = \pi^3 \lambda^2 P_i/4$. We note from (63) that $\lim_{\gamma \to P_w} T_1'(\gamma) > 0$, $\lim_{\gamma \to \infty} T_1'(\gamma) < 0$ and $T_1'(\gamma)$ is a continuous function for $\gamma \geq P_w$. Thus, there is at least one solution of $\gamma$ to the equation $T_1'(\gamma) = 0$ for $\gamma \geq P_w$. We further note that there would be a single solution of $\gamma \geq P_w$ to $T_1'(\gamma) = 0$, which would also be $\gamma^o$, if $T_1(\gamma)$ is a (strictly) quasiconcave function of $\gamma$ for $\gamma \geq P_w$. Thus, in what follows, we prove that $T_1(\gamma)$ is a quasiconcave function of $\gamma$ for $\gamma \geq P_w$.

We prove the quasiconcavity of $T_1(\gamma)$ by the (converse) second-order condition for quasiconcavity. That is, if a function satisfies the condition: at any point with zero slope, the second derivative is non-positive, then the function is quasiconcave.

Taking the second derivative of $T_1(\gamma)$ with respect to $\gamma$, we have

$$T_1''(\gamma) = \frac{\sqrt{B_1}}{\sqrt{\pi}} \left( \frac{B_1 \exp(-\frac{B_1}{\gamma})}{\gamma^{\frac{7}{2}}} - \frac{3 \exp(-\frac{B_1}{\gamma})}{2 \gamma^{\frac{5}{2}}} - \frac{B_1 \exp(-\frac{B_1}{\gamma - P_w})}{(\gamma - P_w)^{\frac{7}{2}}} + \frac{3 \exp(-\frac{B_1}{\gamma - P_w})}{2 (\gamma - P_w)^{\frac{5}{2}}} \right). \tag{64}$$

When $T_1'(\gamma) = 0$, solving for $B_1$ to $T_1' = 0$, we have

$$B_1 = \frac{3\gamma (\gamma - P_w)}{2 P_w} \ln \left( \frac{\gamma}{\gamma - P_w} \right). \tag{65}$$

Substituting (65) into (63), we find

$$\text{sgn} \left( T_1''(\gamma) \right) = \text{sgn} \left( 3 P_w - 3 (2\gamma - P_w) \ln \left( \frac{\gamma}{\gamma - P_w} \right) \right). \tag{66}$$

With the inequality on logarithm $\log(x) \geq 1 - 1/x$, we have

$$3 P_w - 3 (2\gamma - P_w) \ln \left( \frac{\gamma}{\gamma - P_w} \right) \leq \frac{3 P_w (P_w - \gamma)}{\gamma}. \tag{67}$$

Thus, when $T_1'(\gamma) = 0$, we note that $T_1''(\gamma) < 0$ for $\gamma > P_w$, which proves that $T_1(\gamma)$ is a (strictly) quasiconcave function for $\gamma > P_w$. We hence obtain $\gamma^o$ as the solution of $\gamma > P_w$ to $T_1'(\gamma) = 0$, i.e., the solution of $x > P_w$ to (24). Finally, substituting $\gamma^o$ into (62) obtains the average covert probability in (23). This completes the proof.

## Appendix D

## Proof of Theorem 2

Similar to the proof of Theorem 1 given in Appendix B, we here first show the proof that $\eta$ is not affected by $\lambda_I$, and the proof that $\eta$ is not affected by $P_I$ can be easily shown with the similar method. We can rewrite the average covert probability in (31) and the connection outage probability in (33) as

$$\bar{\xi} = \int_0^\infty f_h(x) - \max_{\gamma > P_a d_{aw}^{-\alpha} x} \sum_{k=1}^\infty f_2(k,x) \left( \left( \frac{\lambda_I}{(\gamma - P_a d_{aw}^{-\alpha} x)^{\frac{2}{\alpha}}} \right)^k - \left( \frac{\lambda_I}{(\gamma)^{\frac{2}{\alpha}}} \right)^k \right) \mathrm{d}x \qquad (68)$$

and

$$p_{\text{out}}^{\text{cn}} = \int_0^\infty \sum_{k=1}^\infty f_2(k,x) \left( \frac{\lambda_I}{P_a^{\frac{2}{\alpha}} x^{\frac{2}{\alpha}} d_{ab}^{-2} \left( 2^R - 1 \right)^{-\frac{2}{\alpha}}} \right)^k \mathrm{d}x, \qquad (69)$$

respectively, where

$$f_2(k,x) = \frac{(-1)^{k+1} \Gamma \left( \frac{2k}{\alpha} \right) \sin \left( \frac{2k\pi}{\alpha} \right) \left( \pi \Gamma \left( \frac{\alpha-2}{\alpha} \right) \mathbb{E} \left\{ |h_{ij}|^{\frac{4}{\alpha}} \right\} P_I^{\frac{2}{\alpha}} \right)^k f_h(x)}{k! \pi}. \qquad (70)$$

Following the similar steps given in Appendix B, we can prove that the systems with $\lambda_I = \lambda_{I1}$ and $\lambda_I = u\lambda_{I1}$ have the same covert throughput for any $u > 0$. This completes the proof of Theorem 2.

## APPENDIX E
## PROOF OF PROPOSITION 2

Substituting (35) and (36) into (19), we have

$$\bar{\xi} = \int_0^\infty \left( 1 - \max_{\gamma > P_w} \int_{\gamma - P_w}^\gamma \frac{\pi^{\frac{3}{2}} \lambda_I \sqrt{P_I}}{4x^{\frac{3}{2}}} \exp \left( -\frac{\pi^4 \lambda_I^2 P_I}{16x} \right) \right) \frac{d_{aw}^\alpha}{P_a} \exp \left( -\frac{d_{aw}^\alpha P_w}{P_a} \right) \mathrm{d}P_w. \qquad (71)$$

To derive the average covert probability, we need to determine the optimal threshold of Willie's detector for a given $P_w$, which is given by

$$\gamma^o = \arg \max_{\gamma > P_w} \ T_2(\gamma), \qquad (72)$$

where $T_2(\gamma) = \int_{\gamma - P_w}^\gamma \frac{\pi^{\frac{3}{2}} \lambda_I \sqrt{P_I}}{4x^{\frac{3}{2}}} \exp \left( -\frac{\pi^4 \lambda_I^2 P_I}{16x} \right) = \text{erf} \left( \frac{\pi^2 \lambda_I \sqrt{P_I}}{4\sqrt{\gamma - P_w}} \right) - \text{erf} \left( \frac{\pi^2 \lambda_I \sqrt{P_I}}{4\sqrt{\gamma}} \right)$. The average covert probability can be then rewritten as

$$\bar{\xi} = 1 - T_2(\gamma^o). \qquad (73)$$

Taking the first derivative of $T_2(\gamma)$ with respect to $\gamma$, we have

$$T_2'(\gamma) = \frac{\sqrt{B_2}}{\sqrt{\pi}} \left( \frac{\exp(-\frac{B_2}{\gamma})}{\gamma^{\frac{3}{2}}} - \frac{\exp(-\frac{B_2}{\gamma - P_w})}{(\gamma - P_w)^{\frac{3}{2}}} \right), \qquad (74)$$

where $B_2 = \pi^4 \lambda^2 P_I / 16$. We note that (74) is similar to (63) in Appendix C. Following the similar steps given Appendix E, we can prove that $T_2''(\gamma) < 0$ when $T_2'(\gamma) = 0$ for $\gamma > P_w$.

Hence, $T_2(\gamma)$ is a (strictly) quasiconcave function for $\gamma > P_w$, and $\gamma^o$ is the solution of $\gamma > P_w$ to $T_2'(\gamma) = 0$. Finally, substituting $\gamma^o$ into (73) obtain the average covert probability in (37). This completes the proof.

## Appendix F
## Proof of Corollary 1

In the following, we present the proof that $\eta$ increases as $\lambda_I$ increases. The proof that $\eta$ increases as $P_I$ increases can be shown with the similar method, and hence, is omitted here for brevity.

With the equations of $\bar{\xi} = 1 - \epsilon$ and $p_{\text{out}}^{\text{cn}} = \delta$, $\eta$ is given as the solution of $R$ to

$$\delta = \sum_{k=1}^{\infty} f_1(k) \left( \frac{\lambda_I}{\left( \left[ P_a^* d_{ab}^{-\alpha} \left( 2^R - 1 \right)^{-1} - \sigma_{z_b}^2 \right]^+ \right)^{\frac{2}{\alpha}}} \right)^k, \tag{75}$$

where $f_1(k) = \frac{\Gamma\left( \frac{2k}{\alpha} \right) \left( \pi \Gamma\left( \frac{\alpha-2}{\alpha} \right) P_I^{\frac{2}{\alpha}} \right)^k}{k! \pi} \sin\left( k\pi \left( \frac{\alpha-2}{\alpha} \right) \right)$ and $P_a^*$ is the solution of $P_a$ to

$$\epsilon = \max_{\gamma > P_a d_{aw}^{-\alpha}} \sum_{k=1}^{\infty} f_1(k) \left( \left( \frac{\lambda_I}{(\gamma - P_a d_{aw}^{-\alpha})^{\frac{2}{\alpha}}} \right)^k - \left( \frac{\lambda_I}{(\gamma)^{\frac{2}{\alpha}}} \right)^k \right). \tag{76}$$

Note that the expression for the optimal $P_a$, i.e., $P_a^*$, is the same as that for the case of no AWGN, i.e., the solution of $P_a$ to (55), since $\bar{\xi}$ is not related to the non-zero AWGN.

Denote the optimal $P_a$ when $\lambda_I = \lambda_{I1} > 0$ by $P_{a1}^*$. Denote the covert throughput when $\lambda_I = \lambda_{I1} > 0$ by $\eta_1$. We have

$$\delta = \sum_{k=1}^{\infty} f_1(k) \left( \frac{\lambda_{I1}}{\left( \left[ P_{a1}^* d_{ab}^{-\alpha} \left( 2^{\eta_1} - 1 \right)^{-1} - \sigma_{z_b}^2 \right]^+ \right)^{\frac{2}{\alpha}}} \right)^k. \tag{77}$$

For any $u > 0$, denote the optimal $P_a$ when $\lambda_I = \lambda_{I2} = u\lambda_{I1} > 0$ by $P_{a2}^*$. Denote the covert throughput when $\lambda_I = \lambda_{I2}$ by $\eta_2$. From the analysis on the case of no AWGN, we know that $P_{a2}^* = u^{\frac{\alpha}{2}} P_{a1}^*$. We then have

$$\delta = \sum_{k=1}^{\infty} f_1(k) \left( \frac{\lambda_{I1}}{\left( \left[ \gamma_1^* d_{ab}^{-\alpha} \left( 2^{\eta_2} - 1 \right)^{-1} - u^{-\frac{\alpha}{2}} \sigma_{z_b}^2 \right]^+ \right)^{\frac{2}{\alpha}}} \right)^k. \tag{78}$$

Since $\alpha \geq 2$, we have $u^{-\frac{\alpha}{2}} < 1$ when $u > 1$. Hence, comparing (77) and (78), we have $\eta_2 > \eta_1$ when $m > 1$. In addition, we have $\lambda_{I2} = u\lambda_{I1} > \lambda_{I1}$ when $u > 1$. Therefore, $\eta_2 > \eta_1$ when $\lambda_{I2} > \lambda_{I1}$. This completes the proof.